\documentclass[12pt]{article}

\usepackage{jheppub}
\usepackage{amssymb}
\usepackage{amsmath}
\usepackage{amstext}
\usepackage{amsfonts}
\usepackage{bbold}
\usepackage{bm}
\usepackage{graphics}
\usepackage{mathtools}
\usepackage{lmodern}
\usepackage{graphicx}
\usepackage{caption}
\usepackage{subcaption}
\usepackage[capitalise]{cleveref}
\usepackage{xcolor}
\usepackage{microtype}
\usepackage{graphicx}
\usepackage{mathrsfs}

\pdfoutput=1

\newcommand{\be}{\begin{equation}}
\newcommand{\ee}{\end{equation}}

\newcommand{\tr}{\text{tr}}

\newcommand{\sfrac}[2]{{\textstyle\frac{#1}{#2}}} 

\begin{document}

\title{Solidity without inhomogeneity:\\[.2cm]Perfectly homogeneous, weakly coupled, UV-complete solids}

\author[a]{Angelo~Esposito}
\affiliation[a]{Theoretical Particle Physics Laboratory (LPTP), Institute of Physics, EPFL, 1015 Lausanne, Switzerland}

\author[b]{Rafael~Krichevsky}
\affiliation[b]{Department of Physics, Center for Theoretical Physics, Columbia University, New York, NY, 10027, USA}

\author[b]{Alberto~Nicolis}

\abstract{Solid-like behavior at low energies and long distances is usually associated with the spontaneous breaking of spatial translations at microscopic scales, as in the case of a lattice of atoms.
We exhibit three quantum field theories that are renormalizable, Poincar\'e invariant, and weakly coupled, and that admit states that on the one hand are perfectly homogeneous down to arbitrarily short scales, and on the other hand have the same infrared dynamics as isotropic solids.
All three examples presented here lead to the same peculiar solid at low energies, featuring very constrained interactions and transverse phonons that always propagate at the speed of light. {In particular, they violate the well known $c_L^2>\frac{4}{3}c_T^2$ bound, thus showing that it is possible to have a healthy renormalizable theory that at low energies exhibits a negative bulk modulus (we discuss how the associated instabilities are absent in the presence of suitable boundary conditions).}
We do not know whether such peculiarities are unavoidable features of large scale solid-like behavior in the absence of short scale inhomogeneities, or whether they simply reflect the limits of our imagination.} 

\maketitle


\section{Introduction}

All solids we know of are inhomogeneous and anisotropic at short distances. For instance, in a crystal atoms arrange themselves in a well ordered lattice structure, which is invariant only under a discrete subgroup of translations and rotations. In fact, the acoustic phonons of a solid can be thought of as the Goldstone bosons associated with the spontaneous breaking of translations~\cite{Leutwyler:1996er}. Through the inverse-Higgs mechanism, they also serve as Goldstone bosons for the spontaneously broken rotations (and boosts)~\cite{Nicolis:2013lma}.

Yet, at large enough distance scales, solids look like homogeneous and, sometimes, isotropic continuous media. These two physical properties---large-scale homogeneity and large-scale isotropy---have different origins, and one is more universal than the other. The more universal one is large-scale homogeneity: at distances much larger than the lattice spacing, the fundamental discrete translational symmetry of the lattice is well approximated by continuous translations. This is akin to other accidental symmetries that arise to lowest order in a low-energy expansion, such as baryon number in the standard model of electroweak interactions.
Contrary to those, however, the ``approximate" continuous translational  symmetry of solids is in fact exact {\em to all orders} in a small gradient expansion. This applies to all the fields that live in the solid, including the phonons themselves. 

To convince oneself that this is indeed the case, it's enough to consider the example of a one-dimensional solid. The unbroken discrete translational symmetry  forces all coefficients in the action or Hamiltonian for the fields living in the solid to be periodic functions of $x$, with period given by the lattice spacing $a$:\footnote{Notice that this statement is independent of whether there is an actual lattice  made up of zero-size points, as it is sometimes assumed, or whether instead the discrete translational symmetry is achieved by a more continuous (but inhomogeneous) mass distribution.}
\begin{align}
g (x+a) = g(x) \, ,
\end{align}
where $g$ is a generic such coefficient. This means that $g$ can be expressed in Fourier {\em series}, or, equivalently, that its
Fourier {\em transform} is a sum of delta functions,
\begin{align}
\tilde g(k) = \sum_{m=-\infty}^{+\infty} g_m \times  (2\pi) \delta \big(k - 2 \pi m/{a} \big)\, .
\end{align}
Now, the small gradient expansion corresponds to Taylor-expanding in $k$ around $k=0$. But all the delta functions above with $m \neq 0$ have a trivial Taylor expansion around $k = 0$. To all orders in $k$, only the $m=0$ survives, and we are left with:
\begin{align}
\tilde g(k) \to g_0 \times  (2\pi) \delta (k) \, ,
\end{align}
which, going back to position space, yields
\begin{align}
g(x) \to g_0 = {\rm const} \, .
\end{align}
This is invariant under {\em continuous} translations.

As an explicit example, consider the simple case of a one-dimensional harmonic chain of point-like atoms with mass $m$, spring constant $\kappa_s$ and lattice spacing $a$ (see e.g.~\cite{srivastava2019physics}). In the long wavelength limit, $ka\ll1$, the Hamiltonian for the system can be written as
\begin{align}
H=\int dx \left\{ \frac{\dot u^2}{2} + \frac{c_s^2}{2} \left[ \left(\frac{\partial u}{\partial x} \right)^2 - \frac{a^2}{3}\left(\frac{\partial^2 u}{\partial x^2}\right)^{\!2} + \dots \right] \right\}\,,
\end{align}
where $u(x)$ is the lattice fluctuation field and $c_s^2=\kappa_s a^2 / m$ the speed of sound, while the dots stand for terms of higher order in $ka\ll 1$. As one can see, the higher-derivative corrections due to finite lattice spacing exhibit continuous translational symmetry.

On the other hand, large-scale isotropy, although still very common, arises only if the solid is made up of many domains with random orientations of the underlying lattice structure, and if one looks at distances large enough so that anisotropic effects average to zero. However, it makes perfect sense to also consider very large, possibly infinite anisotropic lattices, which at distances larger than the lattice spacing can be viewed as homogeneous but still anisotropic continuous media. 

The mechanical deformations of  solids at distances larger than the lattice spacing are well described by effective field theories for the Goldstone bosons associated with the spontaneously broken translations \cite{Leutwyler:1996er, Soper:1976bb, Dubovsky:2005xd}. In recent years, these theories have been applied in a number of contexts, from elasticity theory \cite{Alberte:2018doe, Esposito:2018sdc, Soper:2019gob} to cosmic inflation \cite{Endlich:2012pz, Ricciardone:2016lym, Kang:2015uha}.\footnote{Other approaches to model the spontaneous breaking of spatial translations and rotations include holographic theories (see e.g.~\cite{Vegh:2013sk,Davison:2013jba,Amoretti:2014zha,Alberte:2015isw,Alberte:2016xja,Amoretti:2016bxs,Amoretti:2017axe,Amoretti:2018tzw,Li:2018vrz,Andrade:2018gqk,Amoretti:2019cef}), as well as non-renormalizable field theories with or without Lorentz invariance (see e.g.~\cite{Nitta:2017mgk,Musso:2018wbv,Musso:2019kii}). The corresponding Goldstone modes---the phonons---have instead been identified holographically in~\cite{Esposito:2017qpj,Alberte:2017oqx,Baggioli:2019abx,Ammon:2019apj}. {Note that the models of holographic solids present low-energy dynamics that are far less constrained than those we obtain here, with transverse phonons that can be arbitrarily nonrelativistic and quite general effective Lagrangians. However, the focus of our paper is to identify possible UV-completions for the EFTs of solids.
Their existence is not guaranteed a priori, in the same sense as it is not guaranteed that a given gravity theory in the bulk of AdS is the low-energy limit of a healthy theory of quantum gravity.}} Consistently with our remarks above, in these theories there is no sign of the underlying lattice structures of the solids they model, although there can be residual large scale anisotropic effects \cite{Kang:2015uha}. Notice however that these theories feature {\em two} copies of continuous translations: one is the fundamental one, spontaneously broken by the lattice structure, and thus non-linearly realized by the Goldstone fields; the other is the aforementioned approximate-but-exact-to-all-orders continuous limit of the discrete translational symmetry, unbroken by the lattice, and thus linearly realized by the Goldstone fields.  

Given all of the above, it is then natural to ask: Can one have solidity without inhomogeneity? By which we mean: can one have, in some relativistic quantum field theory, a state  whose low-energy excitations have exactly the same dynamics as those of a solid (the acoustic phonons), but which   is nonetheless perfectly homogeneous (and, possibly, isotropic) down to arbitrarily short distances? And, more ambitiously:  for which solid effective theories that are consistent at low energies can this be done (cf.~\cite{Adams:2006sv})?

Phrased in this way, this is a general quantum field theory question, which it would be interesting to understand and answer at some fundamental level, non-perturbatively. Our modest approach in this paper instead is to look for weakly coupled renormalizable theories that achieve what's being asked. We succeed only partially: we are able to reproduce the low-energy dynamics of a solid with cubic symmetry, but these turn out to be those of a highly relativistic solid (some phonons travel at the speed of light). Moreover, if we tune the parameters of the theory in order to remove anisotropies and end up with an isotropic solid, we are left with a still highly relativistic, isotropic solid with very constrained interactions,  much more constrained that those of the most general isotropic solid. We verify this particular limit in three independent ways, using three different constructions to approach it.

\vspace{1em}

\noindent\emph{Conventions:} We set $c=\hbar=1$ and, unless otherwise specified, we adopt a metric $\eta_{\mu\nu}=\text{diag}(-1,1,1,1)$.


\section{The EFT for solids} \label{sec:eft}

From an effective field theory standpoint (EFT), solids that in three spatial dimensions are homogeneous and isotropic at large distances break Poincar\'e together with an internal Euclidean group down to time translations and a diagonal Euclidean group---i.e.
\begin{align} \label{pattern}
ISO(3,1)\times ISO(3)\to \mathbb{R}_t \times ISO(3) \, .
\end{align}
This symmetry breaking pattern can be implemented by means of three real scalars $\phi^I(x)$---the comoving coordinates of the solid's volume elements---which, at equilibrium, can be aligned with the physical coordinates,
\begin{align} \label{vev}
\big\langle\phi^I(x)\big\rangle=\alpha \, x^I \; ,
\end{align}
where $\alpha$ is an arbitrary constant that measures the level of compression or dilation of the solid. Under the internal $ISO(3)$ group the scalars transform as coordinates,
\begin{align} \label{internal}
\phi^I\to O^I_{\;J}\phi^J+c^I\,,
\end{align}
with $c^I$ an arbitrary constant vector, and $O^I {}_{J}$ an arbitrary constant $SO(3)$ matrix~\cite{Endlich:2012pz,Nicolis:2015sra,Esposito:2017qpj}.

It is important to mention that the value of the constant $\alpha$ is not necessarily determined dynamically. This is because the bulk equations of motion of a solid are obeyed for any value of $\alpha$: $\alpha$ is measure of the level of compression or dilation of the solid. For a finite-volume solid with free boundaries, $\alpha$ will relax to the value corresponding to zero pressure or tension. But nothing prevents us from considering more general configurations with arbitrary values of $\alpha$ {(i.e. excited states)}, corresponding to having an external pressure or tension acting on the solid, or, mathematically, to fixed boundary conditions. {This is closely analogous to what happens to an ordinary gas which, in the absence of external constraints, will expand and eventually relax to a state of zero pressure---the vacuum, in that case.}

The low-energy fluctuations around equilibrium are the Goldstone bosons associated with the spontaneous symmetry breaking pattern above, $\phi^I(x)=\alpha \, (x^I+\pi^I(x))$, which are the (acoustic) phonons of the solid.
The most general effective action for the phonons must then be invariant under the Poincar\'e group and the internal $ISO(3)$. To lowest order in derivatives, the only quantity that is Poincar\'e and shift invariant is $B^{IJ}=\partial_\mu\phi^I\partial^\mu\phi^J$. The three independent $SO(3)$ invariants that can be built out of $B^{IJ}$ can be conveniently parametrized as $X=\tr B$, $Y=\tr(B^2)/X^2$ and $Z=\tr(B^3)/X^3$, so that $X$ is the only quantity that changes when $\alpha$ is changed---$Y$ and $Z$ are invariant under rescalings of our solid \cite{Endlich:2012pz}.

The most general low-energy action for the phonons is then
\begin{align} \label{eq:F}
 S=\int d^4x\,F(X,Y, Z)\,,
\end{align}
where $F$ is an a priori generic function, in one to one correspondence with the equation of state of the solid at hand. The action above describes all possible interactions of the phonons among themselves, with effective couplings given by derivatives of $F$ evaluated at equilibrium.

If the solid in question retains some anisotropies at large distances and is invariant only under a discrete subgroup of rotations, the costruction above still applies, but the matrices $O^I {}_J$ in \eqref{internal} have to be taken in that subgroup, and as a consequence the function $F$ in \eqref{eq:F} will be invariant only under that subgroup, and will be thus be a more general function of $B^{IJ}$ rather than just a function of $X$, $Y$, and $Z$---see e.g.~\cite{Kang:2015uha}.


\section{UV-complete, weakly coupled homogeneous solids}

As anticipated in the Introduction, we now try to construct a weakly coupled renormalizable theory that achieves the symmetry breaking pattern just described without ever involving inhomogeneities. Were we to succeed, this would provide a UV-completion for a solid that is homogeneous down to arbitrarily short distances: a solid without an underlying crystal structure.

Trying to reproduce directly the symmetry breaking pattern \eqref{pattern} is tricky. The reason is the $ISO(3)$ factor on the l.h.s., which should act as an internal symmetry. Being non-compact, it does not admit finite dimensional unitary representations. There are however a number of ways to {\em approximate} the symmetry breaking pattern \eqref{pattern} with arbitrary precision.

The first is to consider a discrete subgroup of $ISO(3)$: in particular, we will keep the translation part of $ISO(3)$  continuous, but we will restrict to the cubic subgroup of $SO(3)$. In this way, we can think of the three continuous translations as acting on the phases of three complex scalars, thus making up $U(1)^3$. Then, the cubic subgroup of rotations simply acts as permutations of these three scalars. One can then check if the parameters of the theory can be tuned in such a way as to make anisotropic effects arbitrarily small.

The second is to consider a solid on a very large sphere, so that the isometry group of flat space---$ISO(3)$---is replaced by that of a 3-sphere---$SO(4)$---, which is compact and thus admits unitary finite-dimensional representations. One can then take a suitable flat-space limit, which corresponds to  zooming in on a patch much smaller than the  radius of the sphere.

Finally, one can keep spacetime flat, but consider $SO(4)$ rather than $ISO(3)$ as internal subgroup. One can then take a suitable ``contraction" of $SO(4)$ that reduces to $ISO(3)$, which is simply the group-theoretic version of the zooming-in procedure mentioned above for the sphere.

We analyze these three possibilities in turn. As we will see, at low energies they all reduce to the same peculiar solid.


\subsection{The cubic solid, and an isotropic limit thereof} \label{sec:cubic}

Consider a theory with three complex scalar fields, $\Phi^I$, with an internal $U(1)^3$ symmetry, acting in the obvious way, and a $\Phi^I \leftrightarrow \Phi^J$ permutation symmetry. The most general Poincar\'e-invariant renormalizable theory that is compatible with the above symmetries is
\begin{align}
  \label{eq:L1}
  \begin{split}
  S &= - \int d^4x \sum_I \left[ \big| \partial\Phi^I \big|^2 - m^2 \big| \Phi^I \big|^2 + \frac{\lambda_1}{2} \big| \Phi^I \big|^4 \right] -\int d^4x \, \frac{\lambda_2}{2}  \sum_{I \neq J} \big| \Phi^I \big|^2 \big| \Phi^J \big|^2 \, ,
  \end{split}
\end{align}
where the sign of the mass term has been chosen for later convenience.

We now look for field configurations that break the $U(1)^3$ symmetry by a nontrivial vev. We write the complex scalars in the polar parametrization:
\begin{align}
  \Phi^I (x) = \rho_I (x) e^{i \phi^I (x)} \; .
\end{align}
The phases are the comoving coordinates of our solid, which shift under the internal $U(1)^3$. Under the permutation symmetry they instead transform as $\phi^I\leftrightarrow\phi^J$, showing that this theory represents a solid with cubic symmetry. In terms of these fields, the action becomes
\begin{align} \label{eq:LUV}
  \begin{split}
  S &= - \int d^4x \sum_I \left[ \big(\partial\rho_I\big)^2 - m^2 \rho_I^2 + \rho_I^2 \big(\partial\phi^I\big)^2 + \frac{\lambda_1}{2}  \sum_I \rho_I^4 \right] - \int d^4x \frac{\lambda_2}{2} \sum_{I \neq J} \rho_I^2 \rho_J^2  \, .
  \end{split}
\end{align}

We now want to integrate out the heavy radial fields $\rho_I$ to obtain a low-energy EFT for the comoving coordinates $\phi^I$.
We find it convenient to define the following objects:
\begin{align} \label{eq:defintions}
  \begin{split}
  \vec{b} \equiv  \begin{pmatrix}  B^{11} - m^2 \\ B^{22} - m^2 \\ B^{33} - m^2 \end{pmatrix} , \quad
  \vec{\rho^2} \equiv \begin{pmatrix} \rho_1^2 \\ \rho_2^2 \\ \rho_3^2 \end{pmatrix}  , \quad {\Lambda} \equiv \begin{pmatrix} \lambda_1 & \lambda_2 & \lambda_2 \\ \lambda_2 & \lambda_1 & \lambda_2 \\ \lambda_2 & \lambda_2 & \lambda_1 \end{pmatrix}  ,
  \end{split}
\end{align}
where again $B^{IJ}=\partial_\mu\phi^I\partial^\mu\phi^J$.
At low energies we can neglect the derivatives of $\rho_I$, and write the action as
\begin{align} \label{eq:Lapprox}
  S \simeq - \int d^4x\left[ \vec{b} \cdot \vec{\rho^2} + \frac{1}{2} \vec{\rho^2} \cdot {\Lambda} \cdot \vec{\rho^2}\right] \, ,
\end{align}
which gives the following equation of motion for the heavy mode
\begin{align} \label{eq:rho2}
  \vec{\rho^2} \simeq - {\Lambda}^{-1} \cdot \vec{b} \, .
\end{align}
Plugging this into Eq.~\eqref{eq:Lapprox} one gets the effective action for the $\phi^I$ fields, which is
\begin{align}  \label{eq:LX}
\begin{split}
  S_{\text{eff}} & = \frac{1}{2}\int d^4x\, \vec{b} \cdot {\Lambda}^{-1} \cdot \vec{b}  \\
  &=- \int d^4x\bigg[m^2 ( \xi - 2\xi') X + \frac{\xi^\prime}{2} X^2 + (\xi + \xi^\prime) \tau^{IJKL} B^{IJ} B^{KL} \bigg]\, ,
\end{split}
\end{align}
where we omitted an unimportant additive constant, and as before $X \equiv {\rm tr} \, B$. We also defined
\begin{align}
  \label{eq:notation}
  \begin{split}
  \xi \equiv \frac{\lambda_1 + \lambda_2}{(\lambda_1 + 2\lambda_2)(\lambda_1 - \lambda_2)}& \, ,  \quad  \xi' \equiv \frac{\lambda_2}{(\lambda_1 + 2\lambda_2)(\lambda_1 - \lambda_2)} \, , \\
  \tau^{IJKL} \equiv&  \sum_{N=1}^3 \delta_N^I \delta_N^J \delta_N^K \delta_N^L \, .
  \end{split}
\end{align}

Now, of course, the action~\eqref{eq:LX} does not describe an isotropic solid, and in fact it does not take the form~\eqref{eq:F}. However, isotropy is only spoiled by the term proportional to $\tau^{IJKL}$, which is not invariant under continuous rotations. To recover isotropy we study a regime where the couplings are such that the last term is negligible, i.e.
\begin{align} \label{eq:limit}
  \lambda_1 + 2\lambda_2 \ll \lambda_1 - \lambda_2 \, ,
\end{align}
since this relation between the couplings implies that
\begin{align}
  \xi + \xi' = \frac{1}{\lambda_1 - \lambda_2} \ll &- \frac{1}{3(\lambda_1 + 2\lambda_2)} \simeq \xi' \sim \frac{1}{\lambda_1 + 2\lambda_2} = \xi - 2\xi' \, .
\end{align}
In this limit, the term involving the anisotropic tensor $\tau^{IJKL}$ is subleading, and the effective action for the comoving coordinates becomes
\begin{align} \label{eq:Leff1}
\begin{split}
  S_{\text{eff}} &= -\int d^4x\, \frac{1}{\lambda_1+2\lambda_2} \left[ m^2 X - \frac{X^2}{6} 
+ O \left( \sfrac{\lambda_1+2\lambda_2}{\lambda_1 - \lambda_2} \right) \right] \, ,
  \end{split}
\end{align}
which is a very special case of \eqref{eq:F}.

One might wonder whether the hierarchy of couplings in Eq.~\eqref{eq:limit} is a natural choice.
On the one hand, in the low-energy effective theory such a hierarchy corresponds to a limit of enhanced symmetry---from cubic rotations to full $SO(3)$---, and should thus be technically natural. On the other hand, as we tried to motivate above, there is no way to implement such an enhanced symmetry in the full UV theory. In particular, the limit \eqref{eq:limit} does not appear to correspond to any new symmetry of our original action \eqref{eq:L1}. We address this puzzle in Appendix~\ref{app:naturalness}.

Going back to \eqref{eq:Leff1}, expanding the fields around their equilibrium configuration, $\phi^I=\alpha (x^I+\pi^I)$, one gets the quadratic action for the phonons:
\begin{align} \label{eq:Sphonons}
\begin{split}
S\supset\frac{\alpha^2(m^2-\alpha^2)}{\lambda_1+2\lambda_2}\int d^4x\bigg[\dot{\vec \pi}^2 - c_L^2\big(\vec\nabla\cdot\vec\pi_L\big)^2 -\big(\partial^J\pi_T^I\big)^2\bigg]\,,
\end{split}
\end{align}
where we split the phonon field in its longitudinal and transverse components, $\vec \pi=\vec \pi_L+\vec \pi_T$, such that $\vec\nabla\cdot\vec\pi_T=\vec\nabla\times\vec\pi_L=0$. Then, the longitudinal and transverse sound speeds are 
\begin{align} \label{eq:speeds}
 c_L^2=1-\frac{2}{3}\frac{\alpha^2}{m^2-\alpha^2}\,,\qquad c_T^2=1\,.
\end{align}
Since by definition $\alpha^2>0$, stability ($c^2_{L,T} > 0$, and positive kinetic energy) and subluminality ($c^2_{L,T} \le 1$) require 
\be
\alpha^2 < \sfrac{3}{5}m^2 \; , \qquad \lambda_1+2\lambda_2>0 \; .
\ee 
In Appendix~\ref{app:stability} we show that these are indeed the conditions for the stability of the fluctuations in the full theory \eqref{eq:L1} as well.

The value of the longitudinal and transverse sound speeds in Eq.~\eqref{eq:speeds} do not respect the standard $c_L^2 > \sfrac{4}{3}c_T^2$ bound~\cite{landau1989theory}. This, however, is not an issue. As we show in Appendix~\ref{app:instability}, the instability related to violating such a bound only concerns finite-volume solids whose boundaries are left free, and can never be triggered by localized perturbations in an infinitely extended solid. Our analysis applies to an infinite solid or, equivalently, to the bulk dynamics of a solid of finite volume but whose boundary conditions are fixed. This is similar to the point we raised in the paragraph after Eq.~\eqref{internal}.

{It is clear that the solid modeled here is far from ordinary. First of all, the violation of the $c_L^2>\frac{4}{3}c_T^2$ bound, as also discussed in Appendix~\ref{app:instability}, is due to the presence of negative bulk modulus. Indeed, in the EFT of solids the Lagrangian density is nothing but minus the energy density~\cite{Endlich:2012pz}. Comparing Eq.~\eqref{eq:Sphonons} with standard elasticity theory at zero temperature~\cite{landau1989theory}, one easily finds that the shear modulus, $\mu$, and the bulk modulus, $K$, are given in terms of the parameters of the UV theory by
\begin{align}
\mu \propto \frac{2\alpha^2\left(m^2-\alpha^2\right)}{\lambda_1+2\lambda_2}\,, \quad K \propto - \frac{2\alpha^2\left(m^2+\alpha^2\right)}{3(\lambda_1+2\lambda_2)}\,,
\end{align}
up to a common positive proportionality constant which depends on the normalization of the phonon field. Given that stability implies $\lambda_1+2\lambda_2>0$, the bulk modulus is negative. As a by-product of our analysis, we thus see that there are Poincar\'e invariant, renormalizabile, perturbative theories with $K<0$, which is then a priori allowed by quantum field theory.}

A second interesting, {but related}, point is that the transverse sound speed is always luminal. This is due to the fact that the effective theory we obtained here only depends on the trace of $B^{IJ}$---see Eq.~\eqref{eq:Leff1}. Indeed, deviations from $c_T^2=1$ would be due to the dependence of the action on $Y$ and $Z$~\cite{Endlich:2012pz,Esposito:2017qpj}. Moreover, this is not a peculiarity of the isotropic limit \eqref{eq:Leff1}: one can check that already for the more general case \eqref{eq:LX}, transverse phonons propagating along $\hat x$, $\hat y$, or $\hat z$ always move at the speed of light. So, with this construction we can only reproduce the dynamics of a highly relativistic solid.

To facilitate comparison with what follows, we notice that rescaling the fields and defining $\lambda\equiv(\lambda_1+2\lambda_2)/6$, $\lambda ' \equiv (\lambda_1 - \lambda_2)$, the effective action \eqref{eq:Leff1} becomes\footnote{{As usual, the phonon EFT will break at energies higher than some UV energy scale $\Lambda_{\rm UV}$.  In the present theory, this is of the order of the mass of the lightest radial modes that have been integrated out, $\Lambda_\text{UV}\sim m$. Such an EFT cutoff has, in general, no relation with the scale of inhomogeneity, even more so here, where the system is homogeneous at all distances. This is contrary to what happens for solids with an underlying lattice, where $\Lambda_\text{UV}$ is of the same order as the inverse lattice spacing (times the phonon speed).}}
\begin{align} \label{eq:finalSeff}
S_\text{eff}=-\int d^4x\,\left[\frac{1}{2}X-\frac{\lambda}{4m^4}X^2 + O \left( \lambda^2/\lambda' \right) \right]\,.
\end{align}


\subsection{Flat limit of a solid on a sphere}

Another possibility is to study a solid living on a 3-sphere of radius $r$.\footnote{The  theories presented in this section and in the following one are heavily inspired by the results obtained in~\cite{Esposito:2017qpj}.} Note that a solid living on a sphere (like a thin spherical shell) is not a spherical solid (like a marble). Taking the large radius limit---i.e.~looking at a patch of size much smaller than the radius of the sphere---one can recover the effective theory for a solid in flat space. 

The spatial part of the symmetry breaking pattern of a solid living on a spherical surface is now $SO(4)\times SO(4) \to SO(4)$, since $SO(4)$ is the isometry group of a 3-sphere. We then consider a theory involving a real scalar multiplet $\vec \Phi$ in the fundamental representation of the internal $SO(4)$. The most general Lorentz invariant renormalizable theory is
\begin{align} \label{eq:SPhi}
  S = - \int d^4 x \sqrt{-g} \left[ \frac{1}{2} \big| \partial\vec{\Phi} \big|^2 - \frac{m^2}{2}  \big| \vec{\Phi} \big|^2 + \frac{\lambda}{4}  \big| \vec{\Phi} \big|^4 \right] \, .
\end{align}
We again choose the sign of the mass term so that the vev of  the scalar field spontaneously breaks internal and spatial rotations by picking out a direction. Using standard angular coordinates for $S_3$, the metric in~\eqref{eq:SPhi} is
\begin{align}
  \begin{split}
  ds^2 &= - dt^2 + r^2 \big[ d\theta_1^2 + \sin^2 \theta_1 \big(d\theta_2^2 + \sin^2\theta_2d\theta_3^2\big)\big]\,.
  \end{split}
\end{align}

To implement the symmetry breaking pattern of a solid on a sphere we consider the following background
\begin{align} \label{eq:vev2}
  \big\langle \vec{\Phi} (\theta) \big\rangle = \bar\rho\,\hat r(\theta) = \bar{\rho} \,R(\theta) \cdot\hat{x}_4 \, .
\end{align}
In the expression above $R$ is an $SO(4)$ rotation matrix:
\begin{align}
  R(\theta) &=  e^{\left( \theta_{3} - \frac{\pi}{2} \right) T^{34}} e^{\left( \theta_{2} - \frac{\pi}{2} \right) T^{24}}e^{\left( \theta_1 - \frac{\pi}{2} \right) T^{14} } \, .
\end{align}
with $(T^{AB})_{CD} = \delta^A_D \delta^B_C - \delta^A_C \delta^B_D$ being the $SO(4)$ generators in the fundamental representation (the indices $A,B, \dots$ range from 1 to 4). 
The background configuration~\eqref{eq:vev2} breaks both spatial and internal $SO(4)$ rotations, but not their diagonal combination.
For the angular components of $\vec \Phi$ in field space, it is the spherical analog of Eq.~\eqref{vev}. 

From the equations of motion one finds $\bar{\rho}^2 = (m^2 - 3/r^2)/\lambda$. Since we are ultimately interested in the large $r$ limit, the positivity of this expression for $\bar \rho^2$ is ensured, provided $m^2$ is positive.

The fluctuations around the background can be parametrized by introducing a radial mode and promoting the angles to fields, i.e.
\begin{align}
\vec \Phi(x)=R(\Theta(x))\cdot\vec \rho(x)\,,
\end{align}
with $\vec\rho(x)=\rho(x)\hat x_4$ and
\begin{align}
 \Theta^i(x)\equiv\frac{\pi}{2}-\phi^i(x)\,,
\end{align}
where from now on $i,j,\,\dots\,=1,2,3$, and so that at equilibrium $\Theta^i=\theta_i$.

It is also useful to introduce a covariant derivative $D_\mu=\partial_\mu+R^{-1}\cdot\partial_\mu R$,
so that the action~\eqref{eq:SPhi} becomes
\begin{align}
\begin{split}
 S&=-\int d^4x\sqrt{-g}\bigg[\frac{1}{2}\big| D\vec\rho\,\big|-\frac{m^2}{2}\rho^2+\frac{\lambda}{4}\rho^4\bigg]  \\
 &=-\int d^4x\sqrt{-g}\bigg[ \frac{1}{2}(\partial\rho)^2+\frac{1}{2}\rho^2(\partial R^{-1}\cdot\partial R)_{44}   - \frac{m^2}{2}\rho^2+\frac{\lambda}{4}\rho^4 \bigg]\,, 
 \end{split}
\end{align}
where with the subscript we indicate the entries of the corresponding matrix. At low energies, we can neglect the kinetic term of the radial mode, and solve its equation of motion. This gives $\rho^2\simeq(m^2-(\partial R^{-1}\cdot\partial R)_{44})/\lambda$. The low-energy effective action then reads
\begin{align}
\begin{split}
 S_\text{eff}&=-\int d^4x\sqrt{-g}\bigg[ \frac{m^2}{2\lambda}(\partial R^{-1}\cdot\partial R)_{44}  - \frac{1}{4\lambda}(\partial R^{-1}\cdot\partial R)_{44}^2 \bigg]\,. 
 \end{split}
\end{align}

We can now take the large radius limit, in line with what done in~\cite{Esposito:2017qpj}. To  this end, we focus on a small patch of the sphere around $\theta_i = \frac{\pi}2$, write $\theta_i=\frac{\pi}{2}-\frac{x_i}{r}$, and expand for $x_i\ll r$. Here $x_i$ are the coordinates of the space tangent to the sphere. In this case the metric becomes $g_{\mu\nu}=\eta_{\mu\nu}+O(1/r^2)$, and the fields reduce to
\begin{align}
\phi^i(x)=\frac{1}{r}(x^i+\pi^i(x))\equiv\alpha(x^i+\pi^i(x))\,.
\end{align}
Note that, for a local observer that can only probe a region of space close to the patch, $1/r\equiv\alpha$ plays the role of a free parameter, which can only be determined by boundary conditions.

In this limit we also have $R(\Theta)=\mathbb{1}-\phi^i(x)T^{i4}+O(1/r^2)$, and hence
\begin{align}
 (\partial R^{-1}\cdot\partial R)_{44}=X+O(1/r^3)\,.
\end{align}
Performing the field redefintion $\phi^i\to\sqrt{\lambda}\phi^i/m$, one finds the low-energy effective action for the phonons in the large radius limit:
 \begin{align}
  S_\text{eff}=-\int d^4x\bigg[\frac{1}{2}X-\frac{\lambda}{4m^2}X^2\bigg]+O(1/r^3)\,,
 \end{align}
 which is the same as in Eq.~\eqref{eq:finalSeff}.


\subsection{Group contraction of an $SO(4)$ theory}

Yet another way of writing down a UV-complete theory that induces the symmetry breaking pattern of a homogeneous and isotropic solid is to employ the so-called Wigner-In{\" o}n{\" u} contraction~\cite{Inonu:1993cw}, which allows one to obtain the $ISO(3)$ algebra starting from the $SO(4)$ one. Note that, contrary to what we did in the previous section, we are doing this only for the internal symmetry group. The underlying spacetime is flat.

Let us briefly review how the group contraction works. Separating the $SO(4)$ generators $T^{AB}$ into those that transform as vectors ($T^{i4}$) and those that transform as tensors ($T^{ij}$) under the $SO(3)$ subgroup acting on the first three directions, the complete algebra is given by
\begin{subequations}
\begin{align}
[T^{ij},T^{k\ell}]&=\delta^{ik}T^{j\ell}+\delta^{j\ell}T^{ik} - (k \leftrightarrow \ell)\,, \\
[T^{ij},T^{k4}]&=\delta^{ik}T^{j4}-\delta^{jk}T^{i4}\,, \\
[T^{i4},T^{j4}]&=T^{ij}\,.
\end{align}
\end{subequations}
Rescaling $T^{i4}=\zeta P^i$ and taking the $\zeta\to\infty$ limit, the $SO(4)$ algebra reduces to the $ISO(3)$ one, with $P^i$ and $T^{ij}$ being the generators of shifts and rotations respectively.

Consider now a real multiplet $\vec\Phi$ in the fundamental representation of $SO(4)$. Under a transformation with parameters $\theta_{AB}$ and rescaled generators, it transforms as
\begin{align}
  \Phi^i &\rightarrow \Phi^i + \theta_{ij} \Phi^j - \frac{1}{\zeta} \theta_{i4} \Phi^4 \, , \quad  \Phi^4 \rightarrow \Phi^4 + \frac{1}{\zeta} \theta_{i4} \Phi^i \, .
\end{align}
Let us now rewrite our multiplet as $\Phi^i(x)=\rho(x)\phi^i(x)$ and $\Phi^4(x)=\zeta\rho(x)$, with both $\rho$ and $\phi^i$ independent of $\zeta$. When \noindent$\zeta\to\infty$, one sees that $\rho$ is invariant under $ISO(3)$ while $\vec\phi$ transforms exactly as the solid comoving coordinates, with $\theta_{ij}$ and $\theta_{i4}$ corresponding to the parameters associated respectively with constant rotations and shifts.\footnote{Interestingly, from this viewpoint the fact that translations are nonlinearly realized is due to the spontaneous breaking of $SO(4)$ because of the large vev acquired by $\Phi^4$.}

Our strategy is now the following. We write down a renormalizable theory in flat space for the above multiplet, integrate out the heavy mode, and then take the $\zeta\gg1$ limit. In doing so, we will also take a suitable limit of the parameters of the original theory, namely small mass and coupling. This is done to ensure that the final theory of phonons is non-trivial for large $\zeta$. 
The starting action is once again 
\begin{align} \label{eq:Sinitial}
  S = - \int d^4 x \left[ \frac{1}{2} \big| \partial\vec{\Phi} \big|^2 - \frac{m^2}{2} \big| \vec{\Phi} \big|^2 + \frac{\lambda}{4}  \big| \vec{\Phi} \big|^4 \right] \, ,
\end{align}
but now spacetime is flat from the outset.

We now break $SO(4)$ with a large vev, $\langle \vec\Phi\rangle=\zeta \,\bar \rho \,\hat x_4$. From the action above we find $\bar \rho^2=m^2/(\lambda \zeta^2)$. To keep $\bar \rho = O(\zeta^0)$ one then needs to rescale the parameters of the action so that $m^2/\lambda = O(\zeta^2)$.

A convenient way of parametrizing the full field is
\begin{align}
  \vec{\Phi}(x) = \zeta  \mathcal{O}(x)\cdot\vec\rho(x) \, , \qquad \mathcal{O}(x) \equiv e^{-\frac{\phi^i(x)}{\zeta}T^{i4}} \, ,
\end{align}
where $\vec\rho(x)=\rho(x)\hat x_4$.
Proceeding in a way very similar to the previous section, one finds the action to be
\begin{align}
 S&=-\int d^4x\bigg[ \frac{\zeta^2}{2}(\partial \rho)^2+\frac{\zeta^2}{2}\rho^2(\partial \mathcal{O}^{-1}\cdot\partial \mathcal{O})_{44}   -\frac{\zeta^2m^2}{2}\rho^2+\frac{\lambda \zeta^4}{4}\rho^4\bigg]\,.
\end{align}
After integrating out the radial mode at low energy, the effective action reads
\begin{align}
 S_\text{eff}=-\int d^4x\bigg[\frac{m^2}{2\lambda}(\partial \mathcal{O}^{-1}\cdot\partial \mathcal{O})_{44}-\frac{1}{4\lambda}(\partial \mathcal{O}^{-1}\cdot\partial \mathcal{O})_{44}^2\bigg]\,. \notag
\end{align}

To recover $ISO(3)$ we take the large $\zeta$ limit, which implies
\begin{align}
(\partial \mathcal{O}^{-1}\cdot\partial \mathcal{O})_{44}=\frac{X}{\zeta^2}+O\left(1/\zeta^3\right)\,.
\end{align}
Before finding the final effective action we need to decide how to take the $\zeta\gg 1$ limit of the parameters of the theory. In particular, consistently with what discussed after Eq.~\eqref{eq:Sinitial}, one could choose, for instance, $m^2=\hat m^2$, $\lambda=\hat \lambda/\zeta^2$ or $m^2=\hat m^2\zeta^2$, $\lambda=\hat\lambda$, with both $\hat m^2$, $\hat \lambda = O(\zeta^0)$. However, one can show that in the large $\zeta$ limit, they both lead to an uninsteresting theory of free phonons.
The only choice that produces an interacting theory is $m^2=\hat m^2/\zeta^2$ and $\lambda=\hat\lambda/\zeta^4$.

In this case, again after a field redefinition, the final effective action turns out to be one more time
\begin{align}
 S_\text{eff}=-\int d^4x\bigg[\frac{1}{2}X-\frac{\hat \lambda}{4\hat m^4}X^2\bigg]+O(1/\zeta^2)\,.
\end{align}


\section{Concluding remarks}

{Our paper addresses a conceptual QFT question: is it possible to have a renormalizable, Poincar\'e invariant theory whose low-energy dynamics are those of a solid, but which does not feature inhomogeneities at any scale? We have found that the answer is `yes'.}
In particular, we have exhibited renormalizable weakly coupled quantum field theories that can reproduce the infrared dynamics of solids without relying---like ordinary solids do---on short-scale inhomogeneities. To be precise: our field theories {\em do} break spatial translations, spontaneously, but they do so while preserving a linear combination of those and certain internal symmetry generators, so that there are some unbroken translation-like generators. As a consequence, directly observable quantities such as the energy momentum tensor are exactly invariant under translations, down to arbitrarily short distances (cf.~\cite{Nicolis:2015sra}). In contrast, for an ordinary crystalline solid, all physical quantities are modulated at short distances with the same periodicity as the underlying lattice structure.

The solid-like behavior we are able to reproduce in this way is highly non-generic: phonons have very constrained interactions, and the transverse ones are always ultra-relativistic---they move at the speed of light. It remains to be seen whether such restrictions are unavoidable within this framework, following perhaps from symmetries, locality, and unitarity in a non-trivial way, or whether they are instead a consequence of our considering only weakly coupled scalar field theories. We leave these  questions for future work.

\begin{acknowledgments}
We are grateful to P.~Meade, R.~Penco, F.~Piazza, and R.~Rattazzi for useful discussions. The work by R.K.~and A.N.~has been partially supported by the US Department of Energy grant de-sc0011941. The work done by A.E.~is supported by the Swiss National Science Foundation under contract 200020-169696 and through the National Center of Competence in Research SwissMAP. 
\end{acknowledgments}


\appendix

\section{RG flow and naturalness of the isotropic limit} \label{app:naturalness}

We want to understand how ``natural" the limit in  Eq.~\eqref{eq:limit} is, according to the standard criterion of so-called technical naturalness.
To address this question, let us work in the simplified case of a solid in $2+1$ space-time dimensions. In polar coordinates for field space, the complete action including the radial modes then reads
\begin{align}
 S=-\int d^3x\left[ \sum_I(\partial\rho_I)^2+ \sum_I b_I\rho_I^2 +\frac{1}{2}\sum_{I,J}\Lambda_{IJ}\rho_I^2\rho_J^2\right]\,, \notag
\end{align}
with now $I,J=1,2$ and
\begin{align}
  \begin{split}
  \vec{b} \equiv \begin{pmatrix}  B^{11} - m^2 \\ B^{22} - m^2 \end{pmatrix}, \quad
  {\Lambda} \equiv \begin{pmatrix} \lambda_1 & \lambda_2 \\ \lambda_2 & \lambda_1\end{pmatrix}.
  \end{split}
\end{align}
For our purposes, it is convenient to define $\epsilon\equiv\lambda_1+\lambda_2$ and $\gamma\equiv\lambda_1-\lambda_2$. The limit we are interested in, analogous to Eq.~\eqref{eq:limit}, is $\epsilon\ll \gamma$.

To lowest order in derivatives, the radial modes acquire the following expectation values:
\begin{align} \label{eq:vevrho}
\langle \rho_1 \rangle = \sqrt{\frac{\gamma \tilde{X} - \epsilon \Delta X}{2\epsilon\gamma}} \,, \quad \langle \rho_2 \rangle = \sqrt{\frac{\gamma \tilde{X} + \epsilon \Delta X}{2\epsilon\gamma}} \,.
\end{align}
Here $\tilde{X}\equiv 2m^2-X$ and $\Delta X\equiv B^{11}-B^{22}$, the latter being the only anisotropic operator.

We now introduce  fluctuations around the configurations~\eqref{eq:vevrho}. After diagonalizing the quadratic term, the action reads
\begin{align} \label{eq:S2plus1}
S = & -\frac{1}{\epsilon} \int d^3x \bigg[ -\frac{\tilde X}{4}-\frac{\epsilon}{\gamma} \frac{\Delta X}{4} + \frac{1}{2}\big(\partial\chi_1\big)^2+\frac{1}{2}\frac{\epsilon}{\gamma}\big(\partial\chi_2\big)^2  + \frac{\tilde X}{2}\left(\chi_1^2+\chi_2^2\right) + \frac{1}{2}\sqrt{\frac{\tilde X}{2}} \chi_1^3 \notag \\
& \quad + \frac{1}{16}\chi_1^4 + \frac{1}{2}\left(\frac{\epsilon}{\gamma}\right)^{3/2}\frac{\Delta X}{\sqrt{2\tilde X}} \chi_2^3 + \frac{1}{16}\left(\frac{\epsilon}{\gamma}\right)^2\chi_2^4  + \sqrt{\frac{\tilde X}{2}} \chi_1 \chi_2^2 \\
& \quad  - \frac{1}{2}\sqrt{\frac{\epsilon}{\gamma}}\frac{\Delta X}{\sqrt{2\tilde X}}\chi_1^2\chi_2 - \frac{1}{4}\sqrt{\frac{\epsilon}{\gamma}}\frac{\Delta X}{\tilde X}\chi_1^3\chi_2 + \frac{1}{4}\chi_1^2\chi_2^2  + \frac{1}{4}\left(\frac{\epsilon}{\gamma}\right)^{3/2}\frac{\Delta X}{\tilde X} \chi_1\chi_2^3 + \dots \bigg]\,, \notag
\end{align}
where we have  normalized the fields $\chi_{1,2}$ in order to pull out a factor of $1/\epsilon$ from the action, and for them to have the same mass term. The dots stand for subleading terms: for each interaction, we have only retained the leading contribution in the $\epsilon\ll \gamma$ limit.

Keeping in mind that the $1/\epsilon$ upfront can be ignored for classical computations, there are two things to notice about the above action: $(i)$ the kinetic term for $\chi_2$ (first line) is suppressed by $\epsilon/\gamma$, and $(ii)$ the cubic and quartic self-interactions of $\chi_1$ (second line) are of order unity. The first property implies that $\chi_2$ can be treated as non-dynamical or, in other words, that it is much heavier than $\chi_1$ and it can therefore be integrated out. The second property, instead, implies that one can build all possible tree level amplitudes starting only from the self-interaction of the $\chi_1$, and that these ones will be of order one. It is easy to convince oneself that, instead, the tree level amplitudes involving $\chi_2$ on the internal legs will be subdominant, since they always involve at least one vertex that is suppressed by powers of $\epsilon/\gamma$. 

This means that to integrate out $\chi_2$ at lowest order in $\epsilon/\gamma$ we can simply set it to zero, and obtain the following intermediate EFT, valid for energies much smaller than its mass, $m_2 = \sqrt{\frac \gamma \epsilon} m_1 =  \sqrt{\frac \gamma \epsilon \tilde X}$. The result is
\begin{align}\label{intermediate}
S_\text{eff} & =-\frac{1}{\epsilon}\int d^3x \bigg[ -\frac{\tilde X}{4} + \frac{1}{2}\big(\partial\chi_1\big)^2 + \frac{\tilde X}{2}\chi_1^2  + \frac{1}{2}\sqrt{\frac{\tilde X}{2}} \chi_1^3 + \frac{1}{16}\chi_1^4 + \dots \bigg]\,.
\end{align}
It is easy to check that further integrating out $\chi_1$ yields the (2+1)-dimensional analogue of Eq.~\eqref{eq:Leff1}.

Again, the dots in \eqref{intermediate} stand for terms that are suppressed in the small $\epsilon/\gamma$ limit. Notice that, to zeroth order in this parameter, the intermediate effective action in \eqref{intermediate}  is isotropic: it does not depent on $\Delta X$, which was the only anisotropic combination of Goldstone fields. This stems from the structure of \eqref{eq:S2plus1}, in which 
$\Delta X$ only enters through couplings involving $\chi_2$. Since at low energies and to lowest order in $\epsilon/\gamma$, $\chi_2$ can be set to zero, $\Delta X$ disappears from the action.

So, as far as naturalness of the isotropic limit is concerned, the situation is the following. The UV theory has two dimensionless couplings consistent with cubic symmetry, $\lambda_1$ and $\lambda_2$, or, equivalently, $\epsilon$ and $\gamma$. The small $\epsilon/\gamma$ limit does not correspond to an enhanced $SO(2)$ symmetry of UV theory, and so it is not technically natural in the standard sense. However, as usual with classically marginal couplings, $\epsilon$ and $\gamma$ feature at most a logarithmic dependence on the scale of new physics at high energies, and so the required fine-tuning to ensure $\epsilon/\gamma \ll 1$ is perhaps not too severe. It so happens that $\epsilon/\gamma$ directly controls the ratio of the masses of the two radial modes. So, when  $\epsilon/\gamma$ is very small, one of the radial modes ($\chi_2$) can be integrated out, and one is left with an intermediate EFT for the other radial mode ($\chi_1$) and the angular modes, with an approximate $SO(2)$ symmetry. From this point on---in the RG direction, from high to low energies---there is an approximate enhanced symmetry, which will not be spoiled by quantum corrections within this low-energy EFT. Notice that our small $\epsilon/\gamma$ limit is tied, in the full theory, to one of the radial mode's being much heavier than the other. Such a hierarchy of masses for scalar fields is the quintessential naturalness problem, usually associated with a quadratic dependence of (squared) masses on high energy physics' scales. In our case, as we emphasized, the necessary fine tuning is only logarithmic rather than quadratic, thanks to the direct relationship between masses and dimensionless couplings. In conclusion, how natural our small $\epsilon/\gamma$ limit looks depends on the scale and field parametrization one is looking at: it goes from being logarithmically unnatural (full theory in cartesian parametrization for the fields), to looking quadratically unnatural (full theory in polar parametrization for the fields), to looking technically natural (low-energy EFT below the mass of $\chi_2$).


\section{Stability of the isotropic solid} \label{app:stability}

Here we check what conditions on the parameters of the theory~\eqref{eq:L1} are needed to ensure stability of the fluctuations around the background. On the $\langle \phi^I\rangle =\alpha x^I$ background, the equations of motion imply
\begin{align} \label{rho_I}
 \langle\rho_I^2\rangle=\frac{m^2-\alpha^2}{\lambda_1+2\lambda_2}\qquad \forall \; I \,.
\end{align}
As already mentioned, $\alpha^2>0$ by definition, and therefore positivity of the above expression can be achieved if
\begin{enumerate}
\item $m^2>\alpha^2>0$, $\lambda_1+2\lambda_2>0$: this case corresponds to having spontaneous symmetry breaking even in the absence of the solid background ($\alpha=0$). There is only a limited range of allowed values for $\alpha$: introducing a nonzero $\alpha$ corresponds to pushing the field towards the center of the potential.  The $\alpha=0$ vacuum is perturbatively stable. We will see below that there is a critical value for $\alpha$ ($\alpha= \frac35 m^2$) beyond which the system becomes unstable.

\item $m^2<0$, $\lambda_1+2\lambda_2<0$: This situation corresponds to having a perturbatively stable origin in the $\Phi^I$ field space. In particular, for $\alpha=0$ there is no spontaneous symmetry breaking. However, the potential is not positive definite for large field values, and the solution \eqref{rho_I} corresponds to a perturbatively unstable configuration, which for $\alpha = 0$ is a saddle point or a maximum of the potential.

\item $0<m^2<\alpha^2$, $\lambda_1+2\lambda_2<0$: This case is the most unstable one: even for $\alpha=0$, there is no perturbatively stable vacuum, in the sense that the potential has no minima. We  discard this option.
\end{enumerate}

Let us now study the stability of the background against small fluctuations. Consider first the mass matrix of the radial modes that we integrated out. With our normalization of fields, this is given by
\begin{align}
 M^2 {}_{IJ}=-\frac{1}{2}\frac{\partial^2\mathcal{L}}{\partial\rho_I\partial\rho_J}=2\langle\rho_I^2\rangle\Lambda_{IJ}\,,
\end{align}
where the matrix $\Lambda$ has been defined in Eq.~\eqref{eq:defintions}, and repeated indices are not summed over. 
The eigenvalues $m_i^2$ of $M^2$ are thus proportional to those of $\Lambda$, which are $\epsilon \equiv \lambda_1+2\lambda_2$ and $\gamma = \lambda_1-\lambda_2$ with twofold degeneracy.  We thus have
\begin{align}
m_1^2 = 2(m^2-\alpha^2) \, , \qquad m_{2,3}^2 = 2(m^2-\alpha^2) \frac{\gamma}{\epsilon} \, .
\end{align}
Recall that the isotropic limit corresponds to $\epsilon \ll \gamma$, in which case $m_1^2 \ll m_{2,3}^2$.

A necessary condition for the stability of the configuration \eqref{rho_I} above is the positivity of the mass matrix for the radial modes. This implies $\alpha^2 < m^2$ and $\gamma/\epsilon > 0$, which isolates case 1 above as the only viable option. 

However, this is not the end of the story. The reason is that in the presence of our nontrivial field configuration for the phases $\phi^I$, there are gradient-energy mixings between the radial modes' fluctuations and those of the angular modes, and these could destabilize the system: even though in the original theory all kinetic and gradient energies are positive definite, once we expand about a background with nontrivial gradients, in principle there can  exist excitations that lower the gradient energy (see e.g.~\cite{Creminelli:2019kjy} for an analysis of this phenomenon in a simpler model). 

With this in mind, we analyze  the quadratic Lagrangian for radial excitations, $\rho^I(x) \equiv \langle \rho_I \rangle + h^I(x)$, and angular ones,
$\phi^I (x)\equiv \alpha x^I +  \varphi^I (x) / \langle \rho_I \rangle$, where the normalization of the latter is chosen for later convenience.
Working in $(t, \vec k \,)$ space, after straightforward algebra from \eqref{eq:LUV} we get
\begin{align}
{\cal L } \to \big| \dot {\vec h} \, \big |^2  + \big |\dot {\vec \varphi} \, \big |^2 - \left( 
\begin{array}{c}
{\vec h} \\ {\vec \varphi}
\end{array}
\right)^\dagger \cdot K \cdot \left( \begin{array}{c}
{\vec h} \\ {\vec \varphi}
\end{array}\right) \, ,
\end{align}
where the $6\times6$ matrix $K$ is defined in terms of $3\times3$ blocks as
\begin{align}
K \equiv  \left( \begin{array}{cc}
{k} ^2 \mathbb{1}+ M^2 &  2 i \alpha \, \vec k \cdot \vec{ \mathbb{1}} \\
- 2 i \alpha \, \vec k \cdot \vec {\mathbb{1 }} & {k} ^2 \mathbb{1 }
\end{array} \right) \, ,
\end{align}
and the notation $\vec k \cdot \vec{ \mathbb{1}}$ is shorthand for ${\rm diag}(k_1,k_2,k_3)$.

Stability of our background configuration at finite $\vec k$ corresponds to positivity of $K$, that is, to its eigenvalues being positive. The general six dimensional problem is quite complicated to analyze. However, we can first simplify it somewhat by focusing on the determinant of $K$: if for some values of $\alpha$ this becomes negative, certainly at least one eigenvalue must have become negative. 

To compute the determinant of $K$, we use standard  results for block matrices, in particular 
\begin{align} \label{blocks}
\det\left( \begin{array}{cc}
A &  B \\
C & D 
\end{array} \right) = \det D \times \det (A-B\cdot D^{-1} \cdot C)\,,
\end{align}
We thus get
\begin{align} \label{determinant}
\det K = k^6 \det \big(M^2 + {k} ^2 \mathbb{1} - 4 \alpha^2  (\hat k \cdot \vec{ \mathbb{1}} \,)^2 \big) \, , 
\end{align}
where, as usual, $\hat k \equiv \vec k /k$. 

Notice that, so far, we have made no approximations, and in the above expressions $\vec k $ is generic. However, we expect instabilities, if present at all, to be there only at low enough $\vec k$'s: in the original theory all gradient energies are positive-definite, which means that if one only considers perturbations with high enough momenta such that the potential and the solid background can be neglected---spontaneous symmetry breaking is an infrared phenomenon---{\em any} background configuration will be stable against those perturbations.

This is evident in the r.h.s.~of \eqref{determinant}: $k^2 \mathbb{1}$ appears as a positive-definite correction to $M^2$ inside the determinant. On the other hand, the last matrix only depends on the direction of $\vec k$---not on its magnitude---and thus survives at arbitrarily low momenta. Since it appears as a negative-definite correction to $M^2$, it can in principle trigger an instability.

We are thus led to consider the positivity conditions for 
\begin{align}
 \det \big(M^2  - 4 \alpha^2  N \big) \, , \quad  N \equiv (\hat k \cdot \vec{ \mathbb{1}} \,)^2 = {\rm diag}(\hat k_1^2, \hat k_2^2, \hat k_3^2) \, .
\end{align}
Notice that, regardless of the direction of $\hat k$, the matrix $N$ obeys
\begin{align}
\tr \, N =1 \, , \qquad 0 \le N_{ij} \le 1  \, .
\end{align}
Now we can use the fact that, in the isotropic limit, there is a hierarchy between the eigenvalues of $M^2$,
\begin{align}
m_{2}^2 = m_3^2 \gg m_1^2 \, .
\end{align}
In this case, to leading order in $m^2_1$ and $N$ we get
\begin{align}
\det \big(M^2  - 4 \alpha^2  N \big)  \simeq m_2^2 \, m_3^2 \, (m_1^2 - 4 \alpha^2 N_{11}) \, ,
\end{align}
where $N_{11}$ is the 1-1 entry of $N$ in the basis of eigenvectors of $M^2$. The approximate expression above can be derived, for instance, by applying again Eq.~\eqref{blocks} in such a basis.

In terms of the original basis, the normalized eigenvector of $M^2$ corresponding to $m_1^2$ is $\frac{1}{\sqrt{3}}(1,1,1)$, and so
\begin{align}
N_{11} = \sfrac13 (\hat k_1^2 + \hat k_2^2 + \hat k_3^2) = \sfrac13 \, .
\end{align}

The system thus develops a low-$\vec k$ instability when the combination
\begin{align}
m_1^2 - \sfrac{4}{3} \alpha^2
\end{align}
becomes negative. This corresponds to $\alpha^2$ exceeding $\frac35 m^2$, precisely as we concluded in sect.~\ref{sec:cubic} through a purely low-energy analysis. 


\section{An instability for $c_L^2<\sfrac{4}{3}c_T^2$?} \label{app:instability}

We now show that the instability arising when $c_L^2<\sfrac{4}{3}c_T^2$  \cite{landau1989theory} can only be induced by modes that diverge at spatial infinity and, therefore, can never be triggered by  localized perturbations in the bulk of a solid, or in a solid whose boundary conditions are kept fixed (e.g.~through an external pressure).

We will first briefly review the derivation of such a bound, following~\cite{landau1989theory}. Consider inducing in the solid a deformation with strain tensor $u_{ij}$ which, in terms of our Goldstone field is given at linear order by $u_{ij} \propto \frac{1}{2}(\partial_i\pi_j+\partial_j\pi_i)$. The corresponding variation of the free energy density of the solid, $f$, is
\begin{align} \label{eq:f}
f=f_0 + \mu \,  \big(u_{ij}-\sfrac{1}{3}\delta_{ij}u_{kk}\big)^2 + \sfrac{1}{2}K \, u_{kk}^2 + O\big(u^3\big) \,,
\end{align}
where $f_0$ is the free energy density at equilibrium, and $\mu$ and $K$ are known as the shear and bulk moduli. To ensure positivity of the variation of the free energy (so that a deformation cannot lower it arbitrarily) one typically demands that both $\mu$ and $K$  be positive.

The longitudinal and transverse sound speeds in terms of the moduli are given by~\cite{landau1989theory}
\begin{align}
c_L^2 = \frac{3K+4\mu}{3\rho_m}\,, \qquad c_T^2 = \frac{\mu}{\rho_m}\,,
\end{align}
where $\rho_m$ is the mass density of the solid.\footnote{For simplicity and for consistency with \cite{landau1989theory}, we are considering a non-relativistic solid, but the derivation works equally well for relativistic solids upon minor notational modifications.} Their ratio is therefore $\frac{c_L^2}{c_T^2}=\frac{4}{3}+\frac{K}{\mu}$ and, for positive $K/\mu$, is bounded from below by $4/3$. In order to evade this bound one thus needs $K/\mu<0$. Let's consider the $\mu > 0$, $K <0 $ case, {which is the case relevant for us (see Section~\ref{sec:cubic})}: the free energy~\eqref{eq:f} is unstable against pure-trace deformations, i.e.~deformations such that $u_{ij}-\frac{1}{3}\delta_{ij}u_{kk}=0$. In terms of the Goldstones, this reads
\begin{align}
\partial_i\pi_j+\partial_j\pi_i = \frac{2}{3}\delta_{ij}\vec{\nabla}\cdot\vec{\pi}\,.
\end{align}
This is nothing but the conformal Killing equation in 3D Euclidean space---see e.g.~\cite{francesco2012conformal}---whose four independent solutions correspond to infinitesimal dilations and special conformal transformations:
\begin{align}
\pi_i=a \,  x_i \; , \quad \text{ and } \quad \pi_i=b_i \, \vec{x}^{\,2}-2\big(\vec{b}  \cdot\vec{x}\big) \, x_i\,,
\end{align}
with $a$ and $\vec b$ free parameters.

The configurations under which the solid would be unstable are therefore always divergent at spatial infinity and are not associated with localized perturbations. In other words, the $c_L^2>\frac{4}{3}c_T^2$ bound comes from boundary effects, which are irrelevant for infinite solids or in the bulk of a solid with fixed boundary conditions.


\bibliographystyle{JHEP}
\bibliography{biblio}

\end{document}